\documentstyle[prl,aps,twocolumn,psfig]{revtex}
\begin{document}
\draft
\def\ds{\displaystyle}
\newcommand{\ibs}{\mbox{\boldmath $s$}}
\newcommand{\ibl}{\mbox{\boldmath $l$}}
\newcommand{\ibj}{\mbox{\boldmath $j$}}
\newcommand{\ibS}{\mbox{\boldmath $S$}}
\newcommand{\ibL}{\mbox{\boldmath $L$}}
\newcommand{\ibJ}{\mbox{\boldmath $J$}}
\newcommand{\ibsig}{\mbox{\boldmath $\sigma$}}
\newcommand{\ibr}{\mbox{\boldmath $r$}}
\newcommand{\ibR}{\mbox{\boldmath $R$}}
\newcommand{\ibT}{\mbox{\boldmath $T$}}
\newcommand{\ibf}{\mbox{\boldmath $f$}}
\newcommand{\ibn}{\mbox{\boldmath $n$}}
\title{
Orbital Localization and Delocalization Effects
in the U 5{\ibf}$^{\bf 2}$ Configuration: \\
Impurity Problem}

\author{
Mikito Koga\footnote{
present address: Department of Physics, University of California
Davis, CA 95616 U.S.A.
},
Wenjian Liu, Michael Dolg and Peter Fulde
}

\address{
Max-Planck-Institut f{\"u}r Physik Komplexer Systeme,
N{\"o}thnitzer Str. 38, 01187 Dresden, Germany
}

\maketitle

\begin{abstract}
Anderson models, based on quantum chemical studies of the molecule
of U(C$_8$H$_8$)$_2$, are applied to investigate the problem of an U
impurity in a metal.
The special point here is that the U $5f$-orbitals are divided into
two subsets:
an almost completely localized set and a considerably delocalized one.
Due to the crystal field, both localized and delocalized U $5f$-orbitals
affect the low-energy physics.
A numerical renormalization group study shows that every fixed point
is characterized by a residual local spin and a phase shift.
The latter changes between 0 and $\pi/2$, which indicates the competition
between two different fixed points.
Such a competition between the different local spins at the fixed points
reflects itself in the impurity magnetic susceptibility at high temperatures.
These different features cannot be obtained if the special characters of U
$5f$-orbitals are neglected.
\end{abstract}

\pacs{ }

\narrowtext
In the field of heavy-fermion systems, the research area
has been extended from rare-earth systems (mainly Ce based compounds)
to actinide ones (mainly U).
{}From the numerous studies of Ce systems \cite{Fulde88}, we have
learned that the on-site physics (Kondo problem) is essential
for understanding the formation of heavy quasiparticle bands.
In the case of a Ce$^{3+}$ ion, for which the $4f^1$ configuration is most
stable, a singlet state may be formed at the Ce site via the hybridization
between the conduction band and the f-orbitals.
This helps us to understand one possible origin of heavy quasiparticles
in lattice systems, although other mechanisms have been identified
in different materials such as Yb$_4$As$_3$ \cite{Fulde95} and
Nd$_{2-x}$Ce$_x$CuO$_4$ \cite{Fulde93}.
\par

On the other hand, in the case of an U, we cannot easily extend
the on-site physics to the lattice case because of the complicated
atomic structure.
In real materials, U has a $5f^2$ or $5f^3$ configuration.
In order to understand the atomic physics of the U site,
we must take into account the Hund's coupling as well as
direct Coulomb interaction.
As for Ce, the effect of the crystal field is
important for U;
however, it plays a different role here.
The crystal field deforms the degenerate $5f$-orbitals and
brings anisotropy to the hybridizations between the
deformed orbitals and conduction electrons.
Kusunose and Miyake studied how the Hund's coupling affects
the Kondo effect in such impurity systems \cite{Kusu97}.
However, they confined themselves to two-fold degenerate orbitals
in the U $5f^2$ configuration:
one of them shows a hybridization, while the other is
completely localized.
Their Kondo model, to which the numerical-renormalization-group (NRG) method
\cite{Wilson75,Kri80,Sakai92} was applied, leads to the result that the effect
of Hund's coupling is irrelevant and the fixed point is of the
strong coupling type.
The irrelevant coupling between a local spin and conduction
electrons is ferromagnetic near the fixed point.
Such anisotropic systems were also studied by others in the
limit of strong spin-orbit coupling, where the two-channel
Kondo effect was discussed \cite{Cox87,Koga95}.
Although we have to learn the atomic structure of U in metals to
present a more realistic model, it has not been specified yet on
the experimental side.
However, some information which
is useful to solid state physics can be extracted from
accurate quantum chemical calculations of some molecules with relatively
large ligands, 
e.g., Ce(C$_8$H$_8$)$_2$ \cite{Neu89} and U(C$_8$H$_8$)$_2$ \cite{Liu97}.
The ground state of Ce(C$_8$H$_8$)$_2$ 
was found to be an open-shell singlet ($^1A_{1g}$)\cite{Neu89},
which is analogous to a Kondo singlet of magnetic impurities in metals.
A splitting between the lowest singlet and excited triplet states corresponds
to a low-energy scale for heavy fermion systems in Ce cases.
According to recent results based on the multiconfiguration
self-consistent field (MCSCF) calculations of U(C$_8$H$_8$)$_2$\cite{Liu97},
the U $5f$-orbitals can be divided into two subsets.
Two of the seven U 5f orbitals, which have the same symmetry 
($e_{2u}$ in D$_{8h}$ point group) as the ligand HOMO
and thus strongly hybridize with the ligands, are significantly
delocalized ($\sim 90 \%$ $f$-character).
The other five 5f-orbitals have different symmetries
($e_{1u}$, $e_{3u}$, $a_{2u}$) from that of the ligand HOMO, and
are thus almost completely localized ($> 99 \%$ $f$-character).
This extent of delocalized character is not found for lanthanocenes such as
Nd(C$_8$H$_8$)$_2$, where all the $4f$-orbitals are almost completely
localized.
First, we design a model Hamiltonian
to reproduce the low-lying states of U(C$_8$H$_8$)$_2$ \cite{Liu97}
in order to understand the important aspects.
\par

Table~I (a) shows a few low-lying states of U(C$_8$H$_8$)$_2$ taken from
ref.~\cite{Liu97}.
The low-lying states can be reproduced by using an Anderson model
(\ref{eqn:Model1}) where the different characters of the $5f$-orbitals are
taken into account:
\begin{eqnarray}
& & H = H_f + H_l,
\label{eqn:Model1} \\
& & H_f = \sum_{m \sigma} \varepsilon_m f_{m \sigma}^{\dag} f_{m \sigma}
\mbox{} + \sum_{mm'} {U_{mm'} \over 2}
f_{m \sigma}^{\dag} f_{m' \sigma'}^{\dag} f_{m' \sigma'} f_{m \sigma}
\nonumber \\
& & ~~~~~~~~
\mbox{} + \sum_{mm'} {J_{mm'} \over 2} 
f_{m \sigma}^{\dag} f_{m' \sigma'}^{\dag} f_{m \sigma'} f_{m' \sigma}, \\
& & H_l = V_{\pm 2} (l_{\pm 2, \sigma}^{\dag} f_{\pm 2, \sigma} + {\rm h.c.}),
\end{eqnarray}
Here the energy is measured from the ligand orbital.
$H_f$ represents the energy of U $5f$-electrons;
$H_l$ is the hybridization Hamiltonian between the $5f$ and ligand orbitals.
$f_{m \sigma}^{\dag}$ ($f_{m \sigma}$) and $l_{\pm 2, \sigma}^{\dag}$
($l_{\pm 2, \sigma}$) are the creation (annihilation) operators for
the $5f$ and ligand electrons with spin $\sigma$, respectively.
The orbitals are denoted by the magnetic quantum number $m$
($= 0, \pm 1, \pm2, \pm3$ for $f$).
Only one pair of orbitals with $m = \pm 2$ has a noticeable
hybridization \cite{Liu97}.
The second and third terms in $H_f$ are the direct and exchange
interactions among the $5f$-electrons, respectively.
In this Hamiltonian, every term depends on the orbitals deformed by
the crystal field in the molecule.
Although the spin-orbit coupling, which is neglected here, is
relatively large for heavy atoms like U, it was found to be of minor
importance compared with the Hund's coupling.
As far as we assess the low-lying energy levels, it only yields a
slight modification \cite{Liu97}.

The low-lying states shown in Table~I (a) can be reproduced qualitatively
with our Anderson Hamiltonian where the parameters are derived directly
from quantum chemical calculations.
Since there are too many parameters in the Hamiltonian (\ref{eqn:Model1}),
we consider two kinds of parameter sets.
In both sets, $\varepsilon_0$, $\varepsilon_{\pm 1}$,
and $\varepsilon_{\pm 3}$ are lower than $\varepsilon_{\pm 2}$.
In Set I, a fine anisotropy is introduced to the localized orbital
energies, which results in a small splitting of the low-lying states:
\begin{eqnarray}
& & \varepsilon_0 = -1.205,~\varepsilon_{\pm 1} = -1.200,
\nonumber \\
& & \varepsilon_{\pm 2} = -1.150,~\varepsilon_{\pm 3} = -1.210,
\nonumber \\
& & U_{0,0} = U_{1,1} = U_{3,3} = U_{3,-3} =U_{0,1} = 0.700,
\nonumber \\
& & 
U_{0,3} = U_{1,-1} = U_{1,3} = U_{1,-3} = U_{0,2} = U_{1,2} = U_{1,-2}
\nonumber \\
& & ~~~~~~=U_{2,3} = U_{2,-3} = U_{2,2} = U_{2,-2} = 0.650,
\nonumber \\
& & 2J_{m,m'} = 0.040,~V_{\pm 2} = 0.032,
\label{eqn:Model2}
\end{eqnarray}
where the values are measured in atomic units (au$=27.2$ eV).
The Coulomb couplings satisfy
\begin{equation}
U_{mm'} = U_{m'm} = U_{-m,-m'}.
\end{equation}
In Set II, the localized orbitals are degenerate, while more precise
values derived from ab initio HF calculations
are given to the Coulomb couplings:
\begin{eqnarray}
& & \varepsilon_0 = -1.200,~\varepsilon_{\pm 1} = -1.200,
\nonumber \\
& & \varepsilon_{\pm 2} = -1.150,~\varepsilon_{\pm 3} = -1.200,
\nonumber \\
& & U_{0,0} = U_{1,1} = U_{3,3} = 0.700,~U_{3,-3} = 0.71,~U_{0,1} = 0.690,
\nonumber \\
& & U_{0,3} = U_{1,3} = U_{1,-3} = U_{1,2} = U_{1,-2} = U_{2,2} = 0.650,
\nonumber \\
& & U_{2,3} = U_{2,-3} = U_{2,-2} = 0.640,~U_{1,-1} = 0.660,
\nonumber \\
& & U_{0,2} = 0.630,~2J_{m,m'} = 0.040,~V_{\pm 2} = 0.032.
\label{eqn:Model3}
\end{eqnarray}
If the fine anisotropy is not taken into account, the ground state is still
21-fold degenerate.
This is not the case in Table I (a).
We note that the first and second excited states derived from the Anderson
model are degenerate (Table~I (b)), which are actually split due to $D_{8h}$
symmetry in U(C$_8$H$_8$)$_2$.
\par

We then can present an Anderson model to study an U impurity in metals,
replacing the ligand electrons by conduction electrons.
In order to bring out the orbital effect, we reduce the number
of orbitals:
only one delocalized and one two-fold degenerate localized orbitals
are retained.
The delocalized orbital, which is denoted by $M = 0$, represents the
orbitals with $m = \pm 2$.
In the same way, the localized orbitals, which are denoted by $M = \pm 1$,
represent those with $m = 0, \pm 1, \pm 3$.
Anisotropies in the Coulomb and Hund's couplings (see
({\ref{eqn:Model2}) and (\ref{eqn:Model3})) do not affect the low-energy
physics and do not change the high-energy physics qualitatively.
The most important quantity is the splitting between the localized and
delocalized orbitals.
Then the simplified Anderson Hamiltonian is written as
\begin{eqnarray}
& & H = H_{\rm k} + H_f + H_{\rm hyb},
\label{eqn:Model4} \\
& & H_{\rm k} = \sum_{k \sigma} \varepsilon_k
a_{k,0,\sigma}^{\dag} a_{k,0,\sigma}, \\
& & H_f = \sum_{M \sigma} \varepsilon_M f_{M \sigma}^{\dag} f_{M \sigma}
\mbox{} + {U \over 2} \sum_{MM'}
f_{M \sigma}^{\dag} f_{M' \sigma'}^{\dag} f_{M' \sigma'} f_{M \sigma}
\nonumber \\
& & ~~~~~~~~
\mbox{} +  {J_{\rm H} \over 2} \sum_{MM'}
f_{M \sigma}^{\dag} f_{M' \sigma'}^{\dag} f_{M \sigma'} f_{M' \sigma}, \\
& & H_{\rm hyb} = V_0 (a_{k,0,\sigma}^{\dag} f_{0, \sigma}
\mbox{} + {\rm h.c.}),
\end{eqnarray}
where $H_{\rm k}$ represents the energy of conduction electrons.
$a_{k,0,\sigma}^{\dag}$ ($a_{k,0,\sigma}$) is the creation
(annihilation) operator for the conduction electrons with wave number
$k$, orbital quantum number $M = 0$ and spin $\sigma$.
\par

The Anderson Hamiltonian (\ref{eqn:Model4}) is transformed to a hopping
type of Hamiltonian via the standard procedure of the NRG theory \cite{Kri80}.
The transformed Hamiltonian satisfies a recursion relation of the form
\begin{equation}
H_{N+1} = \Lambda^{1/2} H_N + \sum_{\sigma}
(s_{N,0,\sigma}^{\dag} s_{N+1,0,\sigma} + {\rm h.c.}),
\label{eqn:NRG1}
\end{equation}
and the impurity part is given by
\begin{eqnarray}
& & H_0 = \left[\sum_{M \sigma}
{\tilde \varepsilon}_M f_{M \sigma}^{\dag} f_{M \sigma}
\mbox{} + {{\tilde U} \over 2} \sum_{MM'}
f_{M \sigma}^{\dag} f_{M' \sigma'}^{\dag} f_{M' \sigma'} f_{M \sigma} \right.
\nonumber \\
& & ~~~~~~~~
\mbox{} + {{\tilde J}_{\rm H} \over 2} \sum_{MM'}
f_{M \sigma}^{\dag} f_{M' \sigma'}^{\dag} f_{M \sigma'} f_{M' \sigma}
\nonumber \\
& & ~~~~~~~~\left.
\mbox{} + {\tilde \Gamma}^{1/2}
(s_{0,0,\sigma}^{\dag} f_{0,\sigma} + {\rm h.c.})
\right] \Lambda^{-1/2},
\label{eqn:NRG2}
\end{eqnarray}
where the energies are measured in units of the half width of the
conduction band $D$.
$s_{n,0,\sigma}^{\dag}$ ($s_{n,0,\sigma}$) corresponds to the creation
(annihilation) operator for the conduction electrons.
For the logarithmic discretization parameter, $\Lambda = 2 \sim 3$ is used.
We will fix the strength of hybridization, the Coulomb
and Hund's coupling at ${\tilde \Gamma} = 0.100$, ${\tilde U} = 0.650$ and
${\tilde J}_{\rm H} =0.020$, respectively, throughout the remaining part
of this letter.
Although these values are not connected directly with those given in
(\ref{eqn:Model2}) and (\ref{eqn:Model3}),
the ratio of Coulomb and Hund's couplings is based on the analysis of the
model (\ref{eqn:Model1}).
For practical reasons, the hybridization is assumed to be relatively high,
which should not change the low-energy physics as compared with a smaller
hybridization.
\par

As a result of our NRG calculation, several types of fixed points are
obtained, which depend sensitively on the energy splitting between localized
and delocalized orbitals.
Every fixed point can be characterized by a residual local spin (in Fig.~1)
and a phase shift (in Fig.~2).
The latter will be explained later.
Figure~1 shows the size of the local spin at the fixed points depending on
the choice of parameters.
The $f^2$ configuration is found to be more stable than any other one within
the parameter range
$-2({\tilde U} - {\tilde J}_{\rm H}) = -1.26 < {\tilde \varepsilon}_0
< -({\tilde U}-{\tilde J}_{\rm H}) = -0.63$, provided
${\tilde \varepsilon}_0 \simeq {\tilde \varepsilon}_{\pm 1}$.
It is most stable around a point
${\tilde \varepsilon}_0 = {\tilde \varepsilon}_{\pm 1} = -0.945$.
When both orbital energies are far from this point, charge fluctuations
increase between $f^2$ and $f^1$ or $f^2$ and $f^3$.
When ${\tilde \varepsilon}_0 > -0.63$, $f^1$ is more stable than $f^2$,
while in the region ${\tilde \varepsilon}_0 < -1.26$, we find that $f^3$
is more stable than $f^2$.
\par

We can reproduce the excitations near the fixed point by using the following
effective Hamiltonian:
\begin{eqnarray}
& & H_N = \Lambda^{(N-1)/2} \left[\sum_{n=0}^{N-1}
\Lambda^{-n/2} \sum_{\sigma}
(s_{n \sigma}^{* \dag} s_{n+1,\sigma}^* + {\rm h.c.}) \right.
\nonumber \\
& & ~~~~~~\left.
\mbox{} + \varepsilon^* \sum_{\sigma} s_{0 \sigma}^{* \dag} s_{0 \sigma}^*
\mbox{} - w(N) \ibS \cdot \sum_{\sigma \sigma'}
s_{0 \sigma}^{* \dag} (\ibsig)_{\sigma \sigma'} s_{0 \sigma'}^* \right],
\label{eqn:NRG3}
\end{eqnarray}
with a ferromagnetic exchange interaction between the residual
local spin and the quasiparticles defined at the fixed point.
The positive exchange coupling $w(N)$ decreases as $N$ increases.
This type of Hamiltonian was studied by Kusunose and Miyake in the
case of the underscreened Kondo effect \cite{Kusu97}.
Their discussion was restricted not only to the Kondo regime where
charge fluctuations are eliminated but also to the special case
of our model where ${\tilde \varepsilon}_0 = {\tilde \varepsilon}_{\pm 1}$
is satisfied.
In general, the second term in (\ref{eqn:NRG3}) is marginal, while $w(N)$
vanishes at the fixed point.
The excitations from the ground state can be described by a single-particle
Hamiltonian
\begin{equation}
H_N^* = \sum_l [\eta_l^+ g_l^{\dag} g_l + \eta_l^- h_l^{\dag} h_l],
\end{equation}
where $\eta_l^+$ ($\eta_l^-$) is the $l$-th excitation energy of
a particle (hole).
The latter is given by 
\begin{equation}
\eta_l^{\pm} = \left\{
\begin{array}{ll}
\Lambda^{l-1 \mp \delta / \pi} & (N:{\rm odd}) \\
\Lambda^{l-1/2 \mp \delta / \pi} & (N: {\rm even})
\end{array}
, \right.
\label{eqn:NRG4}
\end{equation}
where $1 \ll l \ll N$.
The phase shift $\delta$ depends on $\varepsilon^*$, namely,
${\tilde \varepsilon}_0$ and ${\tilde \varepsilon}_{\pm 1}$.
As shown in Fig.~2, $\delta$ changes monotonously as far as
the residual local spin keeps its size.
This behavior depends only on ${\tilde \varepsilon}_0$ when
$\Gamma$, $U$ and $J_{\rm H}$ are fixed.
In other words, the localized orbitals ($M = \pm 1$) are only relevant to
the size of the local spin at the fixed point.
Usually, $\delta = 0$ implies that the local spin decouples from the
conduction electrons, while $\delta = \pi/2$ means that the size of the local
spin shrinks from $S$ to $S - 1/2$ due to a spin compensation.
The change of phase shift between 0 and $\pi/2$ indicates a competition
between the two fixed points \cite{Koga96}.
The maximum and minimum of $\delta$ depend on
$\Delta = {\tilde \varepsilon}_{\pm 1} - {\tilde \varepsilon}_0$,
which changes the boundaries of the phase diagram in Fig.~1.
In the vicinity of the boundary in Fig.~1, such competition is
observed clearly in the impurity magnetic susceptibility $\chi_{\rm imp}$
at high temperatures.
In Fig.~3, a shoulder appears in $T\chi_{\rm imp}$ within the range
$0.001 < T/D < 0.01$ ($T$ is temperature).
At lower temperatures, the susceptibility shows
a Curie law:
$T \chi_{\rm imp} = 2/3$ for $S = 1$ at the fixed point, while
$T \chi_{\rm imp} \simeq 1/4$ when $S = 1/2$.
\par

Throughout this letter, much attention is paid to the important
individual characters of U $5f^2$ orbitals:
two of the seven orbitals are delocalized and the others are
almost completely localized.
Several low-lying states of U(C$_8$H$_8$)$_2$ can be reproduced
qualitatively by using a realistic Anderson model including
such U $5f$-characters.
Our NRG study of the impurity problem based on a simplified Anderson model
shows that every fixed point is characterized by a residual local spin and
a phase shift, which depend sensitively on the splitting between
the localized and delocalized orbitals.
The change of phase shift between $\delta = 0$ and $\delta = \pi/2$
is due to a competition between two fixed points corresponding to
the two limiting cases.
For the $5f^2$ configuration, as a result of competition between
$S = 1/2$ and $S = 1$ at the fixed points, a shoulder can be observed in
$T \chi_{\rm imp}$ at high temperatures.
Our results in this letter could be extended to infinite dimensional
lattices \cite{Geor96} as was done for the Anderson model in the absence
of orbital degeneracy \cite{Shimi95}.
\par


\begin{table}
\begin{displaymath}
\begin{array}{lllll} \hline
({\rm a})~{\rm ref.}~\protect\cite{Liu97} & &~~~~~&
({\rm b})~{\rm this~work} & \\
 \hline
{\rm state} & {\rm term} & & {\rm state} & {\rm term} \\
(D_{8h}) & {\rm energy} & & (z{\rm -axis}) & {\rm energy} \\
 \hline \\
{}^3E_{3g} & 0.000 & & f_0^1 f_{\pm 3}^1 & 0.000 \\
 \\
{}^3B_{1g},{}^3B_{2g}~~~~ & 0.066
 & & f_{+1}^1 f_{+3}^1, f_{-1}^1 f_{-3}^1~~~~ & 0.101 \\
 \\
{}^3E_{2g} & 0.376
 & & f_{+1}^1 f_{-3}^1, f_{-1}^1 f_{+3}^1 & 0.101 \\
 \\
{}^3A_{2g} & 0.392 & & f_{+1}^1 f_{-1}^1 & 0.416 \\
 \\ \hline
\end{array}
\end{displaymath}
\caption{
Low-lying states in U(C$_8$H$_8$)$_2$.
The energies (eV) are measured from the ground state.
For the data (this work), we use the parameters given in
(\protect\ref{eqn:Model3}).
}
\end{table}
\bigskip
\begin{figure}
\caption{
Phase diagram of the residual local spins $S$ at the fixed points:
$\Delta = {\tilde \varepsilon}_{\pm 1} - {\tilde \varepsilon}_0$
(${\tilde \varepsilon}_0 < 0$) is the splitting energy between the localized
and delocalized orbitals.
}
\end{figure}

\begin{figure}
\caption{
Dependence of the phase shift $\delta$ on the delocalized orbital
energy ${\tilde \varepsilon}_0$.
Here ${\tilde \varepsilon}_{\pm 1} = {\tilde \varepsilon}_0 < 0$ is assumed.
$\delta$ changes abruptly when ${\tilde \varepsilon}_0$ crosses the boundary
of the phase diagram shown in Fig.~1:
in this case, ${\tilde \varepsilon}_0 = -1.02$.
}
\end{figure}

\begin{figure}
\caption{
Temperature dependence of the impurity magnetic
susceptibility.
The localized orbital energy is fixed at
${\tilde \varepsilon}_{\pm 1} = -0.945$.
}
\end{figure}
\end{document}